\documentclass{PoS}

\title{QCD thermodynamics with two flavours of Wilson fermions on large
lattices}

\ShortTitle{QCD thermodynamics with two flavours of Wilson fermions on large
lattices}

\author{\speaker{Bastian B. Brandt} and Anthony Francis \\
        Institut f\"ur Kernphysik, Johannes Gutenberg-Universit\"at Mainz, \\
        Johann Joachim Becher-Weg 45, 55099 Mainz, Germany\\
        E-mail: \email{brandt@kph.uni-mainz.de}}

\author{Harvey B. Meyer and Hartmut Wittig \\
        PRISMA Cluster of Excellence, \\
        Institut f\"ur Kernphysik, Johannes Gutenberg-Universit\"at Mainz, \\
        Johann Joachim Becher-Weg 45, 55099 Mainz, Germany\\
        and \\
        Helmholtz Institut Mainz, Johannes Gutenberg-Universit\"at Mainz, \\
        Johann Joachim Becher-Weg 36, 55099 Mainz, Germany}

\author{Owe Philipsen\\
        Institut f\"ur Theoretische Physik, Goethe-Universit\"at, \\
        Max-von-Laue-Str. 1, 60438 Frankfurt am Main, Germany}

\abstract{We explore the phase diagram of two flavour QCD at vanishing chemical
potential using dynamical $O(a)$-improved Wilson quarks. In the approach to the
chiral limit we use lattices with a temporal extent of $N_t=16$ and spatial
extent $L=32,48$ and 64 to enable the extrapolation to the thermodynamic limit
with small discretisation effects. In addition to an update on the scans at
constant $\kappa$, reported earlier, we present first results from scans along
lines of constant physics at a pion mass of 290~MeV. We probe the transition
using the Polyakov loop and the chiral condensate, as well as spectroscopic
observables such as screening masses.}

\FullConference{The 30th International Symposium on Lattice Field Theory\\
                 June 24 -- 29,  2012\\
                 Cairns, Australia}

\begin{document}

\section{Introduction}

The unresolved question about the order of the phase transition connected to
chiral symmetry restoration in the chiral limit of
two-flavour QCD is the remaining qualitative issue concerning the phase diagram
in the $\{m_{ud},m_s,T\}$ parameter space at zero chemical potential
(see~\cite{maria_plenary} for a review). There are
two possible scenarios~\cite{Pisarski:1983ms,Butti:2003nu}: In the
first scenario the chiral critical line reaches
the $m_{ud}=0$-axis at some tri-critical point $m_s^{\rm tric}$ and the
transition at $N_f=2$ with $m_{ud}=0$ is of second order. Then the
restoration of chiral symmetry belongs to the $SU(2)\times SU(2)\simeq O(4)$
universality class~\cite{Rajagopal:1992qz}. In the second scenario the chiral
critical line never reaches the $m_{ud}=0$ axis and the transition remains first
order for all values of the strange quark mass.
In~\cite{Pisarski:1983ms,Butti:2003nu} it was
shown that the realisation of one of the two scenarios can be linked to the
strength of the anomalous breaking of the $U_A(1)$-symmetry at the transition
point in the chiral limit. The strength of the anomaly can be
probed by looking at correlation functions in scalar and pseudo-scalar channels
and the associated screening masses. Despite a number of recent
studies aiming to extract information about the order of the transition in the
chiral
limit~\cite{Bonati:2009yg,Bornyakov:2009qh,Bornyakov:2011yb,Burger:2011zc} no
study has sufficient control over the systematic
effects.

In these proceedings we present the current status of our study on the
topic, first reported in~\cite{Brandt:2010uw,Brandt:2010bn}, using
non-perturbatively $\mathcal{O}(a)$-improved Wilson fermions at $N_f=2$.
We use lattices with a temporal extent of $N_t=16$ throughout to
suppress discretisation effects, in particular the effect of the explicit
breaking of chiral symmetry introduced by the Wilson term. We aim at
simulations at several quark masses corresponding to zero-temperature pions with
$m_\pi\lesssim300$~MeV with three different volumes for each simulation
points. This will eventually enable us to perform a scaling analysis with
control over the main systematic effects. Here we present results on the
transition temperatures, screening masses and on the strength of the anomalous
breaking of the $U_A(1)$-symmetry.

\section{Setup}
\label{sec:sim}

The simulations are done using non-perturbatively $\mathcal{O}(a)$-improved
Wilson fermions~\cite{Sheikholeslami:1985ij} with two degenerate dynamical
quarks and the configurations are generated using
DD-HMC~\cite{Luscher:2005rx,Luscher:2007es} and MP-HMC~\cite{Marinkovic:2010eg}
algorithms. For more details on the simulation setup see~\cite{Brandt:2010uw}.

To extract information about the order of the transition in the
chiral limit by means of a scaling analysis, it is important to
control and disentangle the different systematic effects that might distort the
scaling properties of the results at finite lattice spacing and volume. Of
particular importance for Wilson fermions in this context is the suppression of
the explicit breaking of chiral symmetry which shows up as a discretisation
effect at finite lattice spacing. Therefore our simulations are done on large
lattices of the size of
$16\times32^3$, $16\times48^3$ and $16\times64^3$. The three different volumes
at each simulation point allow for an extrapolation to infinite volume, and
at this large temporal extent one can expect discretisation effects to be
small (see also~\cite{Philipsen:2008gq}).

We scan in the temperature by varying the bare coupling $\beta$ while $N_t=16$
remains fixed either at fixed hopping parameter $\kappa$ (for heavier quarks)
or along lines of constant physics (for quarks with an associated
$m_\pi\lesssim290$~MeV). We set the scale by using an interpolation of
the zero-temperature results for the Sommer parameter $r_0/a$ in the chiral
limit obtained within the CLS effort in the region of
$5.20\leq\beta\leq5.50$~\cite{Leder:2010kz,Fritzsch:2012wq} and its continuum
result $r_0=0.503\:(10)$~fm obtained in~\cite{Fritzsch:2012wq}. Similarly, the
tuning of the bare parameters to lines of constant renormalised quark masses is
done using the input from results obtained within
CLS~\cite{Fritzsch:2012wq,Brandt:2011jk,Brandt_phd}.

We extract the critical temperature for the deconfinement transition and the
chiral symmetry restoration using the real part of the APE-smeared Polyakov
loop $L_{\rm SM}\equiv\textnormal{Re}\left[\left<L_{\rm SM}\right>\right]$ and
the subtracted chiral condensate~\cite{Bochicchio:1985xa,Giusti:1998wy}
\begin{equation}
\left<\bar{\psi}\psi\right>_{\rm sub} = 2 \: \frac{N_f\:T}{V} \: m_{\rm PCAC}
\: \int d^4 x \left< P(0)\:P(x) \right> \;,
\end{equation}
respectively. Here $P(x)$ is the pseudoscalar density and $m_{\rm PCAC}$ the
PCAC mass. The transition temperatures are defined by the position of the peak
of the associated susceptibilities
\begin{equation}
 \label{susz}
\chi(O) \equiv N_s^3 \: \left( \left<O^2\right> - \left<O\right>^2 \right) ,
\end{equation}
where $O$ is any of the observables above. For the time being all those
observables are unrenormalised. The error analysis has been done using the
bootstrap method with 1000 bins.

\begin{figure}[t]
 \centering
\includegraphics[width=.75\textwidth]{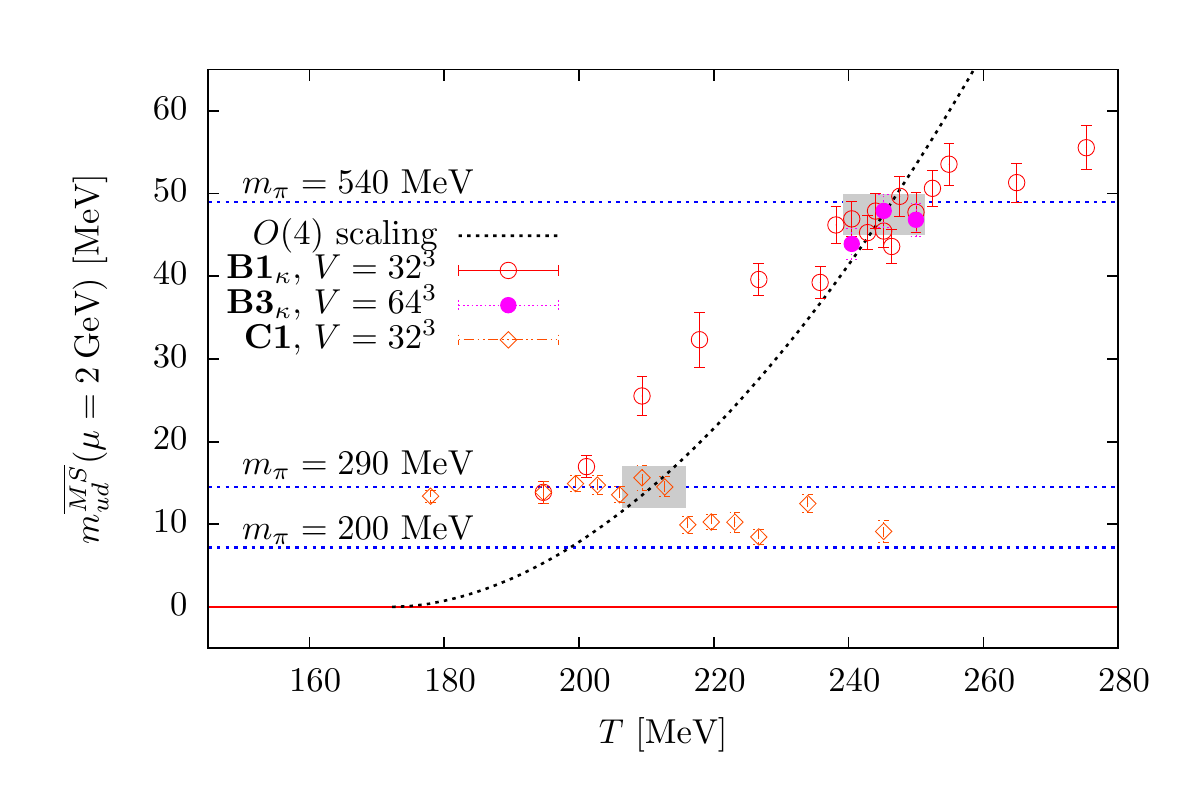}
 \caption{Simulation points in the $\{m_{ud},T\}$-parameter space. The black
dashed line results from naive $O(4)$-scaling for the two transition
temperatures indicated by the grey squares and is only shown to give a
first impression on the possible location of the critical line.}
 \label{fig:sims}
\end{figure}

\section{Current status}
\label{sec:status}

The current set of simulation points in the $\{m_{ud},T\}$-parameter
space~\footnote{Here $m_{ud}$ is the renormalised quark mass in the
$\overline{MS}$-scheme at a renormalisation scale of $\mu=2$~GeV.} is shown in
figure~\ref{fig:sims}. The
set of points consists of two different temperature scans, one at constant
$\kappa$ with two volumes (red and magenta circles) and another one at a line
of constant quark mass of $m_{ud}=14.5$~MeV (orange diamonds). The associated
parameters are listed in table~\ref{tab:sims}.

The results for
scan {\bf B1}$_\kappa$ have in part already been reported
in~\cite{Brandt:2010uw,Brandt:2010bn}. The main results remain unchanged even
with increased statistics and a larger number of simulation points. The
transition temperature in
physical units, listed in table~\ref{tab:ttemps}, changes slightly due to the
updated scale determination reported in~\cite{Fritzsch:2012wq}. Note that the
transition temperature is extracted from the peak position of a Gaussian
fit to the Polyakov loop or condensate susceptibility peak. The
uncertainties are estimated conservatively by the full spread of points included
in the fit. The value given in
table~\ref{tab:ttemps} for scan {\bf B1}$_\kappa$ is extracted from the Polyakov
loop, but the condensate also shows a (weak) peak in its susceptibility at
similar temperature. The additional simulations points from scan {\bf
B3}$_\kappa$ probe the same transition only with twice the volume. The results
are in good agreement with the results for the smaller volume but the limited
number of three simulation points does not allow for a reliable determination 
of the critical temperature. The scan serves as a benchmark for the future
simulations on larger volumes.

\begin{table}[t]
\begin{center}
\begin{tabular}{c|ccccccc}
\hline
\hline
scan & Lattice & DD & $\kappa$ & $m_{ud}$~[MeV] & $T$~[MeV] & $\tau_{U_P}$~[MDU]
&
MDUs \\
\hline
{\bf B1}$_\kappa$ & $16\times32^3$ & $8^4$ & 0.136500 & --- & $190-275$ &
$\sim10$ & $\sim20000$ \\
{\bf B3}$_\kappa$ & $16\times64^3$ & $8\times4^3$ & 0.136500 & --- & $240-250$ &
$\sim46$ & $\sim16000$ \\
\hline
{\bf C1} & $16\times32^3$ & $8^4$ & --- & 14.5 (45) & $175-250$ & $\sim14$ &
$\sim12000$ \\
\hline
\hline
\end{tabular}
\end{center}
\caption{Scans at $N_t=16$ at constant $\kappa$, the ones with subscript
$\kappa$, and at constant renormalised quark mass, {\bf C1}. Listed are the size
of the DD-HMC blocks, DD, the temperature range in MeV, the integrated
autocorrelation time of the plaquette $\tau_{U_P}$ and the
number of molecular dynamics units, MDUs, used in the analysis. The measurements
have been done each 4 MDUs.}
\label{tab:sims}
\end{table}

\begin{table}[b]
\begin{center}
\begin{tabular}{c|ccc}
\hline
\hline
scan & $T_C$~[MeV] & $m_{ud,C}$~[MeV] & $m_{\pi,C}$~[MeV] \\
\hline
{\bf B1}$_\kappa$ and {\bf B3}$_\kappa$ & 245 (7)(6) & 45 ( 2)( 2) & 512
(30)(15)
\\
\hline
{\bf C1} & 211 (5)(3) & 14.5 (20)(13) & 287 (25)(13) \\
\hline
\hline
\end{tabular}
\end{center}
\caption{Estimates for the transition temperatures, quark and pion masses. The
first error bar reflects the uncertainty due to the extraction of the peak
position. The second error denotes the uncertainty due to
scale setting and renormalisation. For scan {\bf B1}$_\kappa$ and {\bf
B3}$_\kappa$ the result for the transition temperature has been extracted from
the Polyakov loop susceptibility, for scan {\bf C2} the value from the
subtracted condensate has been used. The conversion from quark to
pion masses is done using continuum $\chi$PT to NNLO~\cite{Bijnens:1997vq} with
the low energy constants from~\cite{Brandt_phd,new_pub}.}
\label{tab:ttemps}
\end{table}

\begin{figure}[t]
\begin{minipage}[c]{.47\textwidth}
\centering
\noindent
\includegraphics[width=.9\textwidth]{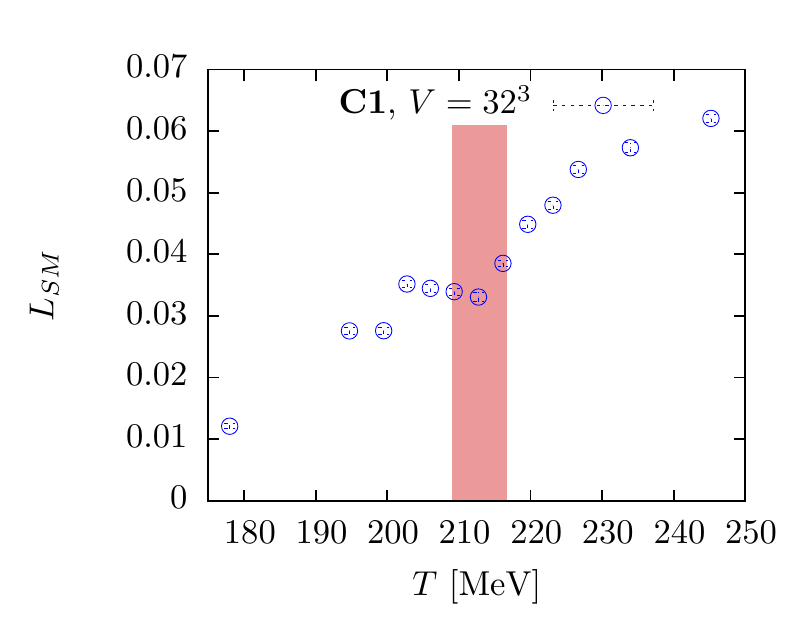}
\newline
\includegraphics[width=.9\textwidth]{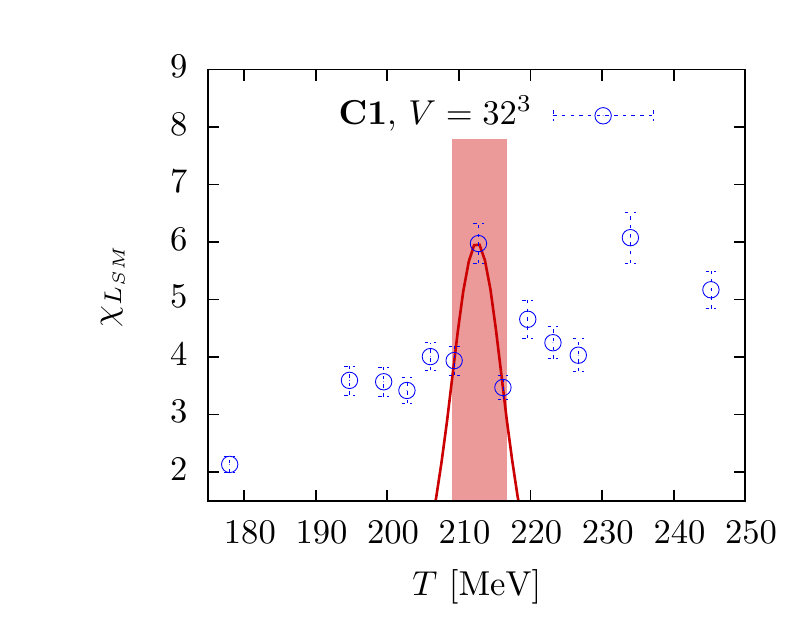}
\end{minipage}
\begin{minipage}[c]{.47\textwidth}
\centering
\noindent
\includegraphics[width=.9\textwidth]{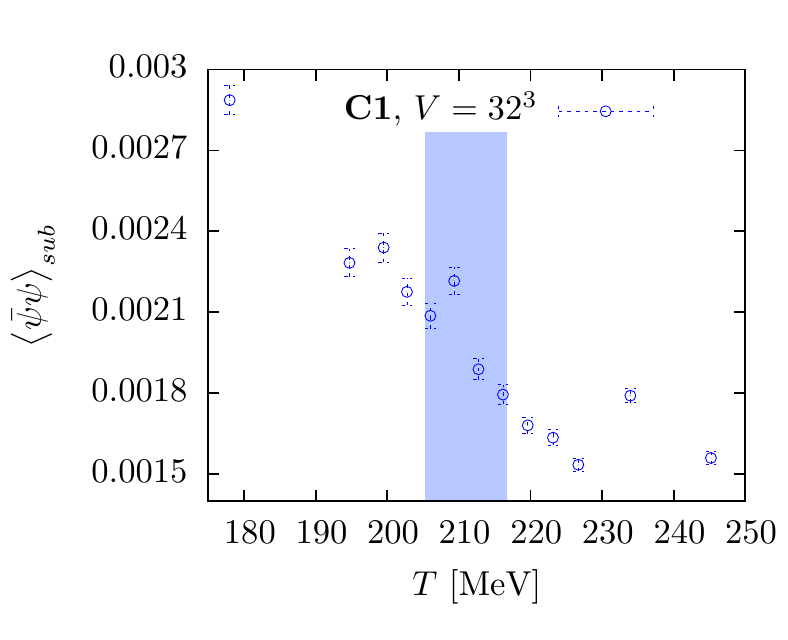}
\newline
\includegraphics[width=.9\textwidth]{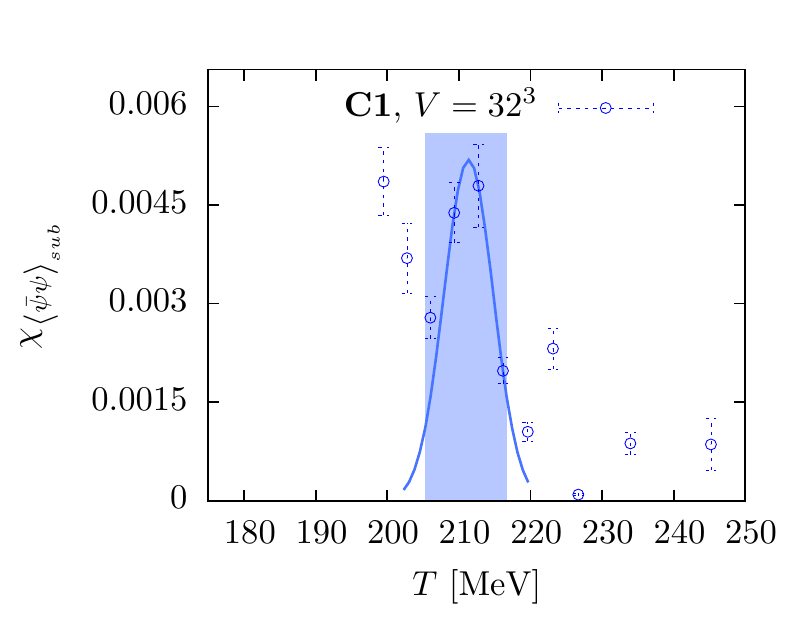}
\end{minipage}
\caption{Results for the smeared Polyakov loop $L_{\rm SM}$ (left) and the
subtracted chiral condensate $\left<\bar{\psi}\psi\right>_{\rm sub}$
(right) and their susceptibilities for scan {\bf C1}. The coloured
areas are the transition regions extracted from the susceptibility. The curves
are the Gaussian fits used to define the transition point.}
 \label{fig:C1}
\end{figure}

Scan {\bf C1} is our first scan along a line of constant quark mass of 14.5~MeV,
which corresponds to a zero-temperature pion mass of about 290~MeV, and
provides first results in the regime relevant for the future scaling analysis.
The results for the smeared Polyakov loop, the subtracted condensate and
their susceptibilities are shown in figure~\ref{fig:C1}. As can be seen from
the plot, both susceptibilities exhibit a peak at a similar temperature. The
result given in table~\ref{tab:sims} is the one extracted from the condensate,
which agrees with the one extracted from the Polyakov loop within errors.

\section{Screening masses}
\label{sec:screening}

Further information about the chiral symmetry restoration pattern in a given
scan can be extracted from the behaviour of screening
masses~\cite{DeTar:1987ar}. In particular, the degeneracy of
pseudoscalar and scalar screening masses signals the restoration of the
anomalously broken $U_A(1)$-symmetry. In this section we focus on scan {\bf C1}
and study screening masses in the pseudoscalar ($P$), scalar ($S$), vector ($V$)
and axial vector ($A$) channels, measured on the
stored configurations (separated by 40~MDUs) with a point source.

\begin{figure}[t]
\centering
\includegraphics[width=.7\textwidth]{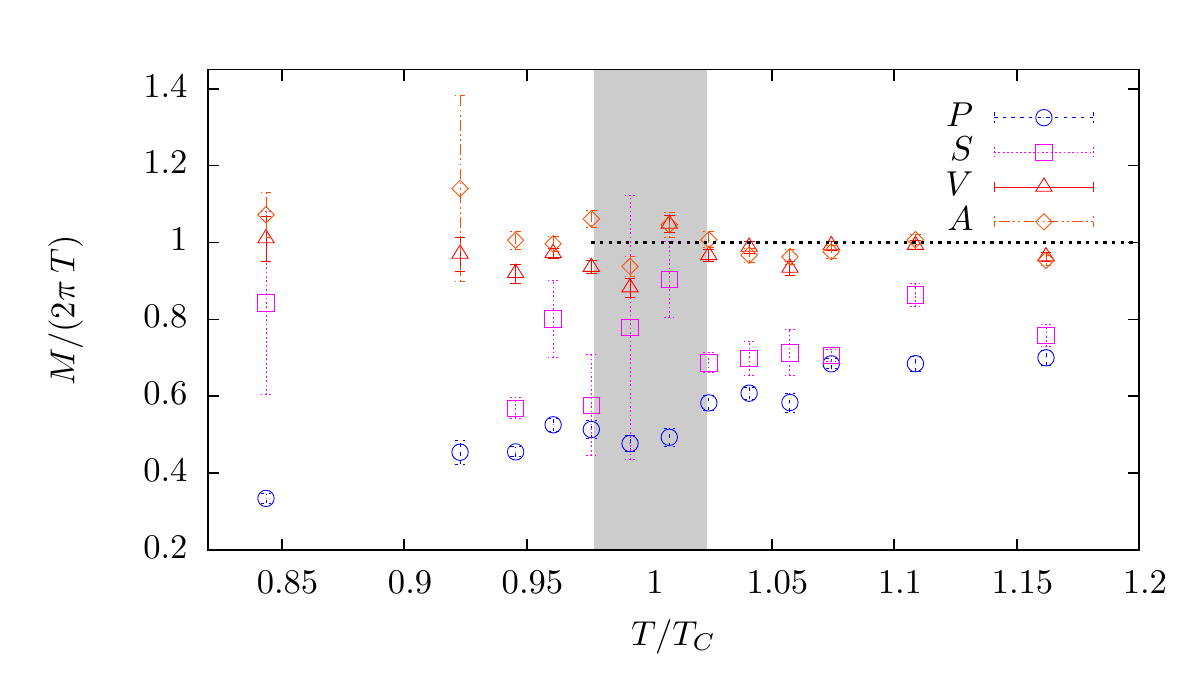}
\caption{Temperature dependence of the screening masses in $P$, $S$, $V$ and
$A$ channels. The dashed line corresponds to the asymptotic value of
$M=2\pi\:T$.}
\label{fig:screening}
\end{figure}

Figure~\ref{fig:screening} illustrates the temperature dependence of the
screening masses. The $x$-axis is normalised to the
critical temperature listed in table~\ref{tab:ttemps}. At $0.84\:T_C$ the
screening masses are mostly in agreement with the expected splitting patterns of
the zero-temperature meson masses. The screening
mass in the pseudoscalar channel, $M_P$, starts from a value
of $M_P/(2\pi\:T)=0.33\:(2)$, which means that at this point $M_P$ is a factor
of 1.27 larger than the zero-temperature pion mass in the scan. Around $T_C$ it
then starts to rise and at $1.16\:T_C$ it is roughly 30~\% smaller than the
asymptotic value $2\pi\:T$. This is in the ballpark of what has been found
for the Wilson action on pure gauge configurations~\cite{Wissel:2005pb}, but
larger than typical results for staggered
fermions~\cite{Cheng:2010fe,Banerjee:2011yd}. Note however, that
regardless of the discretisation finite volume (here $N_s/N_t=2$) and quark mass
effects might still give a sizeable contribution around $T_C$. 
The $V$ and $A$-channels are accidentally
already close to the asymptotic value of $2\pi\:T$ below $T_C$ and
mainly fluctuate around this value in the whole interval. $M_V$ and $M_A$
fluctuate independently below $T_C$ and become degenerate above $T_C$,
consistent with chiral symmetry restoration.

The strength of the anomalous breaking of the chiral $U_A(1)$-symmetry can be
assessed via the splitting between the screening masses in $P$ and $S$ channel.
Below and at $T_C$ the screening mass in the $S$-channel and shows large
fluctuations. Above $T_C$ the signal becomes more stable and $M_S$ moves closer
to $M_P$, signalling a weakening of the breaking of $U_A(1)$. At $1.16\:T_C$
the symmetry is almost restored. This is in qualitative agreement with the
findings from~\cite{bazavov:2012ja} where the symmetry has been found to be
restored at about $1.25\:T_C$.

\section{Conclusions}
\label{sec:conclusions}

This proceedings article contains a summary of the status of our ongoing study
of the QCD deconfinement transition in the chiral limit at $N_f=2$. To date
there are two scans available with $N_t=16$ and pion masses
at the critical point of about 510 and 290~MeV. The critical temperatures are
245 and 211~MeV respectively. The scan at the pion mass of 290~MeV is the first
scan in the region important for a future scaling analysis and is done along a
line of constant renormalised quark mass of 14.5~MeV.
To study the pattern of chiral symmetry restoration,
we calculate mesonic screening masses. The
pseudoscalar screening mass rises from a value close to the zero-temperature
pion mass at $0.84\:T_C$ towards the asymptotic value of $2\pi\:T$. At
$1.16\:T_C$ it differs by roughly 30~\% from this limit. This is in agreement
with the results for screening masses extracted from pure gauge theory with
Wilson fermions~\cite{Wissel:2005pb}, but larger than typical results found in
simulations with staggered
fermions~\cite{Cheng:2010fe,Banerjee:2011yd}.
The results show the expected degeneracy for vector and axial
vector channels around and above $T_C$, signalling chiral symmetry restoration.
At the same time, the $U_A(1)$-symmetry still appears to be broken.
The large lattices used in our study also offer the possibility to study plasma
properties in terms of temporal correlation functions in the vector channel and
the associated spectral function (see~\cite{Ding:2010ga}). A first study has
already been reported at conferences and we refer to our future publication for
the details.

\acknowledgments

The simulations where done on the WILSON cluster at the Institute for
Nuclear Physics of the University of Mainz, on
the FUCHS cluster at the Center for Scientific Computing of the University of
Frankfurt and on JUROPA and JUGENE at FZ Juelich under project number HMZ21. We
are grateful to the institutes for offering these facilities. B.B.
is supported by DFG Grant ME 3622/2-1.

\end{document}